\documentclass[conference]{IEEEtran}
\usepackage{courier}
\usepackage[usenames,dvipsnames]{color} 
\usepackage{color,url,xspace}
\usepackage{amssymb,amsmath}
\usepackage{graphicx}
\usepackage{colordvi}
\usepackage{epsfig}
\usepackage{endnotes}
\usepackage{verbatim}
\usepackage{subfigure}
\usepackage{textcomp}
\usepackage{subfig}
\usepackage{setspace}
\usepackage{xcolor}
\usepackage{amsmath}
\usepackage{hyperref}
\usepackage{algorithm2e}
\usepackage{cite}
\usepackage{epstopdf}
\usepackage{times}
\usepackage{booktabs, multirow}

\usepackage{color}
\usepackage{soul}

\hypersetup{
    pdfauthor={},
    pdfkeywords={},
    pdfpagemode=pagewidth,
    plainpages=false,
    colorlinks,
    urlcolor=blue,
    linkcolor=blue,
    linkbordercolor=red,
    citebordercolor=orange,
    citecolor=blue,
    bookmarksnumbered
}

\begin{document}
\newtheorem{thm}{Theorem}
\newtheorem{problem}{\textbf{Problem}}
\newtheorem{theorem}{\textbf{Theorem}}
\newtheorem{lemma}{\textbf{Lemma}}
\newtheorem{definition}{\textbf{Definition}}
\newtheorem{remark}{\textbf{Remark}}
\newcommand{\argmin}{\operatornamewithlimits{argmin}}
\newcommand{\argmax}{\operatornamewithlimits{argmax}}

\title{Feasibility Study of Stochastic Streaming with \\ 4K UHD Video Traces}
\author{\IEEEauthorblockN{Joongheon Kim$^\dag$ and Eun-Seok Ryu$^\ddag$}
\IEEEauthorblockA{$^{\dag}$Platform Engineering Group, Intel Corporation, Santa Clara, California, USA \\
$^{\ddag}$Department of Computer Engineering, Gachon University, Republic of Korea \\
Emails: joongheon@gmail.com$^{\dag}$, esryu@gachon.ac.kr$^{\ddag}$}
}
\maketitle

\begin{abstract}
This paper performs the feasibility study of stochastic video streaming algorithms with up-to-date 4K ultra-high-definition (UHD) video traces.
In previous work, various stochastic video streaming algorithms were proposed which maximize time-average video streaming quality subject to queue stability based on the information of queue-backlog length.
The performance improvements with the stochastic video streaming algorithms were verified with traditional MPEG test sequences; but there is no study how much the proposed stochastic algorithm is better when we consider up-to-date 4K UHD video traces.
Therefore, this paper evaluates the stochastic streaming algorithms with 4K UHD video traces; and verifies that the stochastic algorithms perform better than queue-independent algorithms, as desired.
\end{abstract}


\IEEEpeerreviewmaketitle

\section{Introduction}\label{sec:intro}
According to the predictions from the Cisco Visual Networking Index (VNI)~\cite{cisco2013}, the summation of all possible forms of video contents will constitute 80\% to 90\% of global data traffic by 2017, and the traffic from mobile and wireless portable devices will exceed the traffic from wired devices by 2016.
Therefore, efficient wireless video streaming algorithms are of the highest importance~\cite{cm2013golrezaei}.

Based on this importance, various types of video streaming algorithms have been investigated; and one of major research directions is \textit{stochastic video streaming}
    which aiming at the time-average video quality maximization subject to video queue/buffer stability~\cite{ton2015kim,arxiv2014kim,mobicom2013kim,asilomar2012bethanabhotla,tcomm2014bethanabhotla,isit2013bethanabhotla}.
In~\cite{ton2015kim,arxiv2014kim,mobicom2013kim}, stochastic video streaming algorithms for device-to-device distributed computing systems are proposed.
In~\cite{ton2015kim}, device-to-device stochastic video streaming with two types of schedulers (centralized vs. distributed) is discussed; and the related performance evaluation with various settings is performed.
In~\cite{asilomar2012bethanabhotla,tcomm2014bethanabhotla,isit2013bethanabhotla}, stochastic video streaming in small cell networks is proposed; and the corresponding theoretical analysis is also presented.

In the two research directions, they discuss about stochastic network optimization applications to adaptive video streaming (i.e., stochastic streaming) which maximizes time-average video streaming quality subject to queue/buffer stability.
If we transmit maximum quality video streams all the time, the streaming quality will be maximized whereas the queue/buffer within the transmitter will be overflowed.
On the other hand, if we transmit minimum quality video streams all the time, the queue/buffer will be stable whereas the streaming quality will be minimized.
Therefore, the proposed stochastic streaming adapts the quality of each video stream depending on current queue-backlog length~\cite{ton2015kim,arxiv2014kim,mobicom2013kim,asilomar2012bethanabhotla,tcomm2014bethanabhotla,isit2013bethanabhotla}.

In~\cite{ton2015kim,arxiv2014kim,mobicom2013kim,asilomar2012bethanabhotla,tcomm2014bethanabhotla,isit2013bethanabhotla},
the used video traces are MPEG test sequences, however the test sequences are not used in current consumer electronics applications.
Therefore, this paper evaluates the stochastic streaming algorithms with up-to-date 4K ultra-high-definition (UHD) video test sequences.
After observing the performance evaluation results with 4K UHD video traces, we can numerically identify how much the novel stochastic streaming algorithm is better than queue-independent non-adaptive video streaming algorithms.

The remainder of this paper is organized as follows:
Section~\ref{sec:2} explains the proposed stochastic video streaming algorithm in~\cite{ton2015kim,arxiv2014kim}.
Section~\ref{sec:3} shows the simulation results with various simulation parameter settings and with the 4K UHD video traces.
Section~\ref{sec:4} concludes this paper and presents future research directions.

\section{Proposed Stochastic Streaming in~\cite{ton2015kim,arxiv2014kim}}\label{sec:2}
\begin{figure}[t!]
	\begin{center}
		\includegraphics[width=0.95\columnwidth]{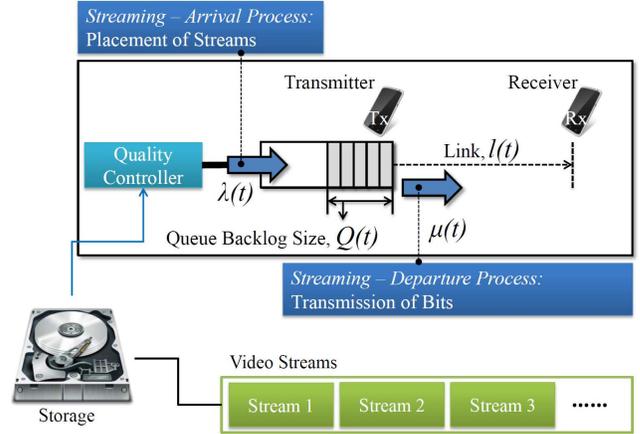}
	\end{center}
	\caption{A stochastic streaming model}
	\label{fig:StochasticStreamingConcept}
\end{figure}

As well-studied in~\cite{ton2015kim,arxiv2014kim} and also shown in Fig.~\ref{fig:StochasticStreamingConcept}, the proposed stochastic video streaming consists of two parts, i.e., (i) placement of streams (i.e., arrival process of the queue/buffer) and (ii) transmission of bits (i.e., departure process of the queue/buffer).

The placement of streams happens in each stream time $t_{s}$ and also the transmission of bits happens in each unit time $t$, respectively. It means both unit time and stream time have different time clock as explained in~\cite{ton2015kim}.
In this paper, stream placement happens when
\begin{equation}
t \textrm{ mod } K = 0
\end{equation}
where $K$ is a positive integer value which is the scaling factor between stream time and unit time.
In addition, stream time can be defined as follows only when $t \textrm{ mod } K = 0$:
\begin{equation}
t_{s} = \left\{
\begin{array}{ll}
  0,           & t    = 0 \\
  \frac{t}{K}, & t \neq 0.
\end{array}
\right.
\end{equation}

In the given system in Fig.~\ref{fig:StochasticStreamingConcept}, the queue dynamics can be formulated as follows:
\begin{equation}
Q(t+1) = \max\left\{Q(t)+\lambda(t)-\mu(t), 0\right\}
\end{equation}
where
    $t\in\{0,1,2,\cdots\}$,
    $Q(t)$ is queue backlog length in unit time $t$,
    $\lambda(t)$ is the arrival process of the queue/buffer (i.e., placement of streams and the details are in Section~\ref{sec:streaming-arrival}),
    and
    $\mu(t)$ is the departure process of the queue/buffer (i.e., transmission of bits and the details are in Section~\ref{sec:streaming-departure}).

\begin{algorithm}[t!]
\SetAlgoLined
\textbf{Parameter setting} \\
$\cdot$ $K$: scaling factor between stream time and unit time \\
$\cdot$ $V$: tradeoff between video quality and queue stability \\
$\cdot$ $M$: set of possible quality modes \\
$\cdot$ $\textsf{BW}$: channel bandwidth of the system \\
$\cdot$ $P^{\textrm{Tx}}$: transmit power \\
$\cdot$ $N_{\textrm{mW}}$: background noise \\
\vspace{3mm}
\textbf{Stochastic video streaming} \\
\While{$t \geq 0$}{
    \If{$t=0$}{
        $\cdot$ $Q[0] \leftarrow 0$\\
    }
    \Else
    {
        // $ t \neq 0$ \\
        $\cdot$ Observe channel state at $t$: $h(t)$ \\
        $\cdot$ Observe current queue-backlog at $t$: $Q(t)$ \\
        \vspace{2mm}
        \textbf{\textit{(1) arrival process calculation}}\\
        $\cdot$ $\lambda(t)\leftarrow 0$ \\
        \If{$t \text{ mod } K = 0$}{
            $\cdot$ $t_{s} \leftarrow \frac{t}{K}$ \\
            $\cdot$ $\mathcal{F}^{*}\leftarrow -\infty$ \\
            \For{$\forall q\left(t_{s}\right)\in M$}{
                \If{$\mathcal{F}^{*} < \Phi\left(q\left(t_{s}\right),t_{s}\right)$ [in Eq. (\ref{eq:phi-def})]}{
                $\cdot$ $\mathcal{F}^{*} \leftarrow \Phi\left(q\left(t_{s}\right),t_{s}\right)$ \\
                $\cdot$ $q^{*}\left(t_{s}\right) \leftarrow q\left(t_{s}\right)$ \\
                }
            }
            $\cdot$ $\lambda(t)\leftarrow \mathbb{B}\left(q^{*}\left(t_{s}\right),t_{s}\right)$ \\
        }
        \vspace{2mm}
        \textbf{\textit{(2) departure process calculation}}\\
        $\cdot$ $\mu(t) \leftarrow \textsf{BW}\cdot\log_{2}\left(1+\frac{P^{\textrm{Tx}}_{\textrm{mW}}\cdot\left\|h(t)\right\|^{2}}{N_{\textrm{mW}}}\right)$ \\
        \vspace{2mm}
        \textbf{\textit{(3) queue update}}\\
        $\cdot$ $Q(t+1) \leftarrow \max\left\{Q(t)+\lambda(t)-\mu(t),0\right\}$
    }
}
\caption{Pseudo-code for stochastic streaming~\cite{ton2015kim}}
\label{algorithm-stochastic}
\end{algorithm}

\begin{figure*}[t!]
	\begin{center}
		\includegraphics[width=1.35\columnwidth]{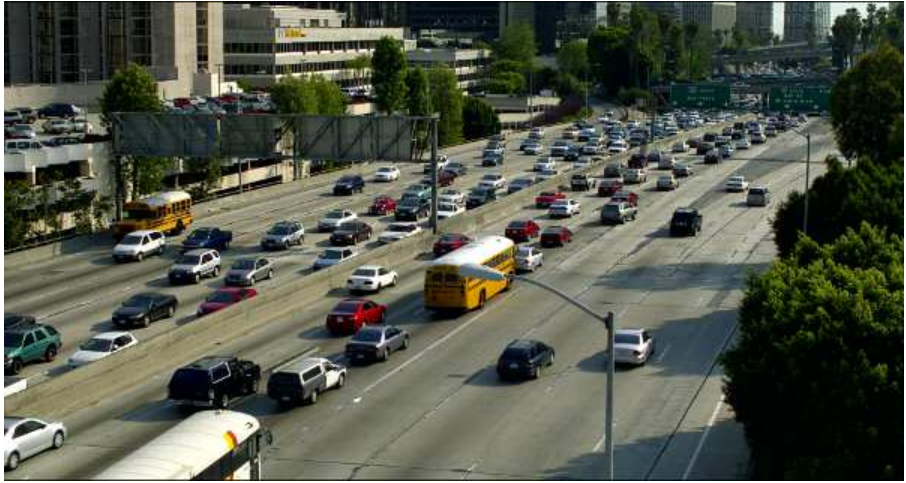}
	\end{center}
	\caption{A sample 4K UHD video frame}
	\label{fig:original}
\end{figure*}

\begin{figure*}[t!]
	\centering
	\subfigure[QP: 22]{
		\includegraphics[width=0.75\columnwidth]{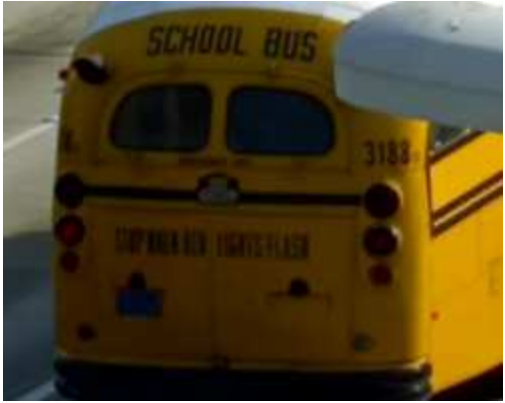}
		\label{fig:SimTraffic10e4v100}
	}
	\subfigure[QP: 27]{
		\includegraphics[width=0.75\columnwidth]{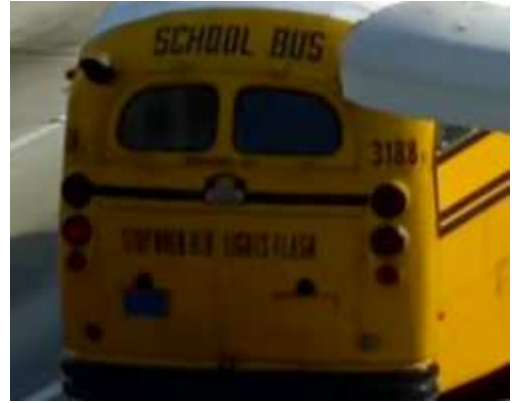}
		\label{fig:SimTraffic14e4v100}
	}
	\subfigure[QP: 32]{
		\includegraphics[width=0.75\columnwidth]{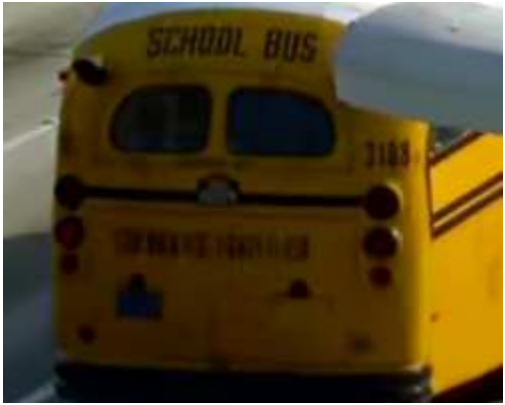}
		\label{fig:SimTraffic10e4v100}
	}
	\subfigure[QP: 37]{
		\includegraphics[width=0.75\columnwidth]{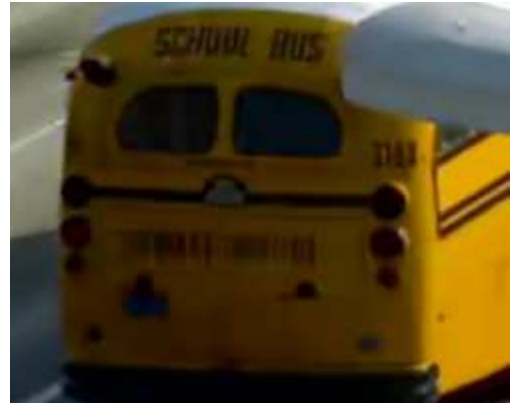}
		\label{fig:SimTraffic14e4v100}
	}
	\caption{Test video sequences \#1: School bus}
	\label{fig:sim1}
\end{figure*}

\subsubsection{Arrival Process (Placement of Streams)}\label{sec:streaming-arrival}
In each stream time slot $t_{s}$, the transmitter of each link places a stream into its transmission queue.
This is the arrival process of the given system, and it is denoted as $\lambda(t)$ in Fig.~\ref{fig:StochasticStreamingConcept}.

In order to dynamically and adaptively select the quality level of the streams by the \textit{Quality Controller} in Fig.~\ref{fig:StochasticStreamingConcept},
we consider stochastic network optimization frameworks for maximizing the total time-average video quality subject to queue stability.

Then, the proposed stochastic optimization problem is given by:
\begin{eqnarray}
\max & & \lim_{t\rightarrow \infty}\frac{1}{t}\sum_{t_{s}=0}^{t-1}\mathbb{E}\left[\mathbb{P}\left(q\left(t_{s}\right),t_{s}\right)\right] \label{eq:obj}\\
\text{subject to} & & \lim_{t\rightarrow \infty}\frac{1}{t}\sum_{t_{s}=0}^{t-1}\mathbb{E}\left[Q\left(q\left(t_{s}\right),t_{s}\right)\right] < \infty 
\label{eq:meanratestable}
\end{eqnarray}
where
    $\mathbb{P}\left(q\left(t_{s}\right),t_{s}\right)$ is the peak-signal-to-noise-ratio (PSNR) of a current stream in stream time $t_{s}$ when the quality mode is $q\left(t_{s}\right)$,
    $Q\left(q\left(t_{s}\right),t_{s}\right)$ is the queue backlog length in stream time $t_{s}$ when the quality mode is $q\left(t_{s}\right)$, and
    (\ref{eq:meanratestable}) stands for the given queue should fulfill queue stability~\cite{book2010neely}.
Note that PSNR is one of representative indices for numerically identifying the quality of video frames~\cite{tbc2013kim}.
In addition, the $\mathbb{P}\left(q\left(t_{s}\right),t_{s}\right)$ and $Q\left(q\left(t_{s}\right),t_{s}\right)$ in Eq. (\ref{eq:obj}) and Eq. (\ref{eq:meanratestable}) can vary depending on quality modes.
If the quality mode is for maximum quality, the $\mathbb{P}\left(q\left(t_{s}\right),t_{s}\right)$ and $Q\left(q\left(t_{s}\right),t_{s}\right)$ in Eq. (\ref{eq:obj}) and Eq. (\ref{eq:meanratestable}) will be maximum by assuming that higher quality streams have the large amounts of bits for more detailed representation of video contents.

As theoretically discussed and proved in~\cite{ton2015kim},
the quality control decision involves choosing our optimal $q^{*}\left(t_{s}\right)$ for the time-average stochastic optimization framework in Eq.~(\ref{eq:obj}) and Eq. (\ref{eq:meanratestable}) as follows:

\begin{figure*}[t!]
	\centering
	\subfigure[QP: 22]{
		\includegraphics[width=0.75\columnwidth]{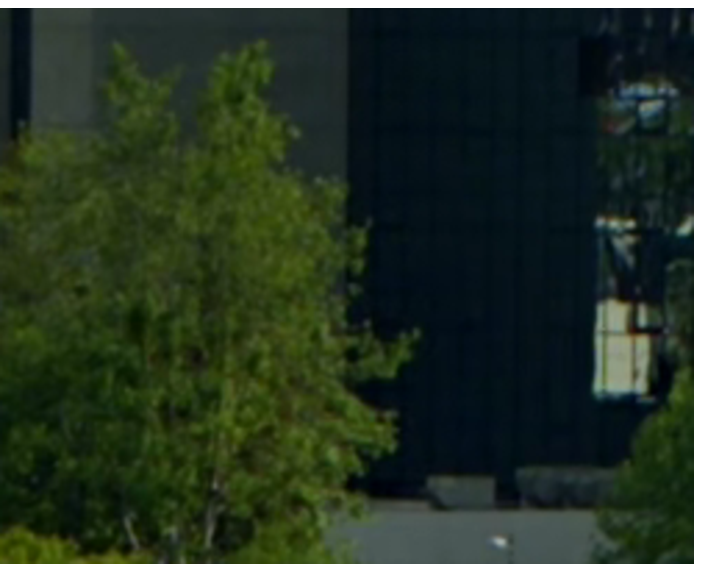}
		\label{fig:SimTraffic10e4v100}
	}
	\subfigure[QP: 27]{
		\includegraphics[width=0.75\columnwidth]{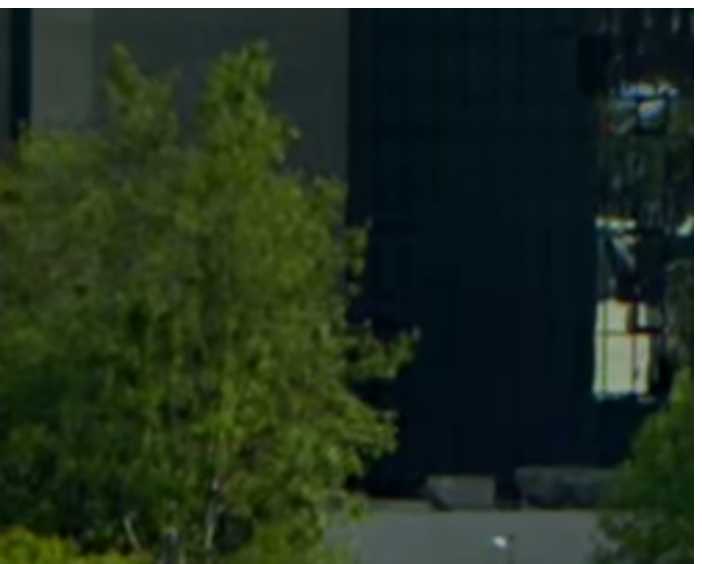}
		\label{fig:SimTraffic14e4v100}
	}
	\subfigure[QP: 32]{
		\includegraphics[width=0.75\columnwidth]{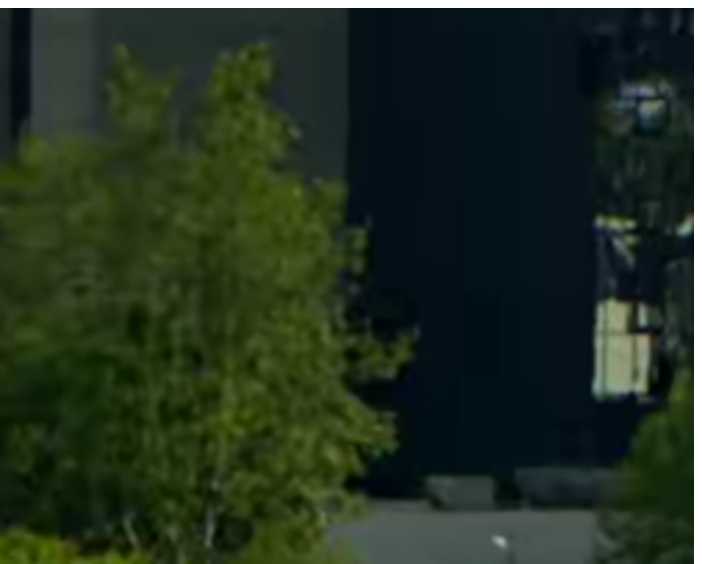}
		\label{fig:SimTraffic10e4v100}
	}
	\subfigure[QP: 37]{
		\includegraphics[width=0.75\columnwidth]{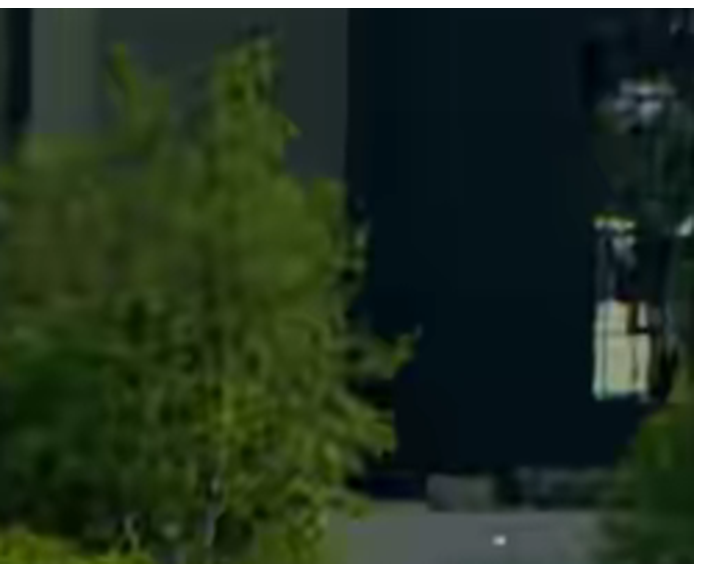}
		\label{fig:SimTraffic14e4v100}
	}
	\caption{Test video sequences \#2: Trees and buildings}
	\label{fig:sim2}
\end{figure*}

\begin{equation}
q^{*}\left(t_{s}\right) \leftarrow \arg\max_{q\left(t_{s}\right)\in M}\Phi\left(q\left(t_{s}\right),t_{s}\right)
\label{eq:final2}
\end{equation}
where
\begin{equation}
\Phi\left(q\left(t_{s}\right),t_{s}\right) \triangleq
\mathbb{P}\left(q\left(t_{s}\right),t_{s}\right) - V \cdot \mathbb{B}\left(q\left(t_{s}\right),t_{s}\right)\cdot Q(t)
\label{eq:phi-def}
\end{equation}
where
    $M$ is the set of possible quality modes,
    $V$ is a tradeoff parameter between quality maximization and queue stability (if this $V$ is small, the optimization framework put more priority on quality maximization, and vice versa), and
    $\mathbb{B}\left(q\left(t_{s}\right),t_{s}\right)$ is the size (i.e., bitrate) of the stream in stream time $t_{s}$ when the quality mode is $q\left(t_{s}\right)$.

Since the placement of streams constitutes the arrival process of the queue,
$\lambda(t)$ can be denoted as follows when the optimal $q^{*}\left(t_{s}\right)$ is determined using Eq. (\ref{eq:final2}).
\begin{equation}
\lambda(t) = \left\{
\begin{array}{ll}
  \mathbb{B}\left(q^{*}\left(t_{s}\right),t_{s}\right), &    t_{s} \text{ mod } K = 0, \\
  0, &  t_{s} \text{ mod } K \neq 0.
\end{array}
\right.
\end{equation}

\subsubsection{Departure Process (Transmission of Bits)}\label{sec:streaming-departure}
As illustrated in Fig.~\ref{fig:StochasticStreamingConcept}, the departure process in the given system is the transmission of bits.
We can transmit bits as much as the wireless channel allows.
In conventional wireless standards (e.g., IEEE 802.11, IEEE 802.15, or 3GPP LTE), modulation and coding scheme (MCS) sets are defined and the corresponding wireless links can transmit bits based on the defined MCS rules.
In this paper, we are not considering specific standard protocols. Therefore, theoretical Shannon's capacity equation is considered and it can be calculated as follows~\cite{book2011molisch}:
\begin{equation}
\mu(t) = \textsf{BW}\cdot\log_{2}\left(1+\frac{P^{\textrm{Tx}}_{\textrm{mW}}\cdot\left\|h(t)\right\|^{2}}{N_{\textrm{mW}}}\right)
 \label{eq:mu}
\end{equation}
where
    $P^{\textrm{Tx}}_{\textrm{mW}}$ stands for the power transmitted by a transmitter to its associated receiver in a milli-Watt scale,
    $h(t)$ stands for the channel gain from the transmitter to its associated receiver at time $t$,
    $N_{\textrm{mW}}$ is a background noise in the system in a milli-Watt scale,
    and
    $\textsf{BW}$ stands for the channel bandwidth of the system.
In (\ref{eq:mu}), the $N_{\textrm{mW}}$ is assumed to be $1$ in this paper.

Finally, the pseudo-code of the proposed stochastic video streaming in~\cite{ton2015kim,arxiv2014kim} is as presented in Algorithm~\ref{algorithm-stochastic}.

\section{Feasibility Study}\label{sec:3}
\subsection{Test Sequence Generation}\label{sec:3-1}
The computing environments and original video information for video trace generation are as follows:
\begin{itemize}
    \item Resolution: 3840-by-2048 (for 4K UHD video)
    \item Frame rate: 30 fps (30 frames per second)
    \item Bit depth: 8 bits
    \item Test sequence name: Traffic (for video standard testing)
    \item Profile name: \texttt{Main}
    \item Intra Period: 32
    \item GOP size: 8
    \item Four different video qualities with QP (quantization Parameters): 22, 27, 32, and 37
    \item Encoder: HM version 15.0 (HEVC standard reference codes)
    \item PC: Intel i7 CPU, Windows7 64bit OS
\end{itemize}

\subsection{4K UHD Video Traces}\label{sec:3-2}

With the computing and parameter settings as presented in Section~\ref{sec:3-1}, 4K UHD test video traces are generated and the representative sample full video frame is as presented in Fig.~\ref{fig:original}.
Two parts of the full video frame are as presented in Fig.~\ref{fig:sim1} and Fig.~\ref{fig:sim2}.
For each part, the compression results are presented in each Fig.~\ref{fig:sim1} and Fig.~\ref{fig:sim2}.
As explained in Section~\ref{sec:3-1}, we have four different quality levels, i.e.,
\begin{equation}
M = \{\textrm{QP=}22, \textrm{QP=}27, \textrm{QP=}32, \textrm{QP=}37\}
\end{equation}
and $q\left(t_{s}\right)\in M$.
For each quality level, visual compression results are presented in Fig.~\ref{fig:sim1} and Fig.~\ref{fig:sim2} for the given two difference parts in the full video frame.

In addition, the PSNR and Bitrate are measured in each quality level for 10 sample streams; and the corresponding measured results are presented in Table~\ref{video-info}.

\begin{table}[t!]%
\caption{Video Trace Information}
\label{video-info}
    \centering %
{\small
	\begin{tabular}{r|r||r|r}
    \toprule[1.0pt]
    Stream \# & Quality \# & PSNR (dB) & Bitrate (Kbps) \\
    & $q\left(t_{s}\right)$ & $\mathbb{P}\left(q\left(t_{s}\right),t_{s}\right)$ & $\mathbb{B}\left(q\left(t_{s}\right),t_{s}\right)$ \\
	\midrule
    1        & 1 (QP: 22) & 41.64	 & 26496 \\
    1        & 2 (QP: 27) & 39.11	 & 10658 \\
    1        & 3 (QP: 32) & 36.61	 & 5073 \\
    1        & 4 (QP: 37) & 34.00	 & 2621 \\
	\midrule
    2        & 1 (QP: 22) & 41.64	 & 26811 \\
    2        & 2 (QP: 27) & 39.07	 & 10811 \\
    2        & 3 (QP: 32) & 36.56	 & 5128 \\
    2        & 4 (QP: 37) & 33.97	 & 2650 \\
	\midrule
    3        & 1 (QP: 22) & 41.60	 & 27888 \\
    3        & 2 (QP: 27) & 39.00	 & 11279 \\
    3        & 3 (QP: 32) & 36.48	 & 5320 \\
    3        & 4 (QP: 37) & 33.91	 & 2721 \\
	\midrule
    4        & 1 (QP: 22) & 41.61	 & 27145 \\
    4        & 2 (QP: 27) & 39.05	 & 10958 \\
    4        & 3 (QP: 32) & 36.53	 & 5193 \\
    4        & 4 (QP: 37) & 33.94	 & 2679 \\
	\midrule
    5        & 1 (QP: 22) & 41.63	 & 26535 \\
    5        & 2 (QP: 27) & 39.08	 & 10710 \\
    5        & 3 (QP: 32) & 36.57	 & 5095 \\
    5        & 4 (QP: 37) & 33.98	 & 2636 \\
	\midrule
    6        & 1 (QP: 22) & 41.60	 & 27630 \\
    6        & 2 (QP: 27) & 39.02	 & 11130 \\
    6        & 3 (QP: 32) & 36.51	 & 5263 \\
    6        & 4 (QP: 37) & 33.94	 & 2703 \\
	\midrule
    7        & 1 (QP: 22) & 41.61	 & 27766 \\
    7        & 2 (QP: 27) & 39.01	 & 11237 \\
    7        & 3 (QP: 32) & 36.49	 & 5303 \\
    7        & 4 (QP: 37) & 33.91	 & 2714 \\
	\midrule
    8        & 1 (QP: 22) & 41.63	 & 26689 \\
    8        & 2 (QP: 27) & 39.10	 & 10765 \\
    8        & 3 (QP: 32) & 36.59	 & 5118 \\
    8        & 4 (QP: 37) & 34.00	 & 2641 \\
	\midrule
    9        & 1 (QP: 22) & 41.62	 & 27083 \\
    9        & 2 (QP: 27) & 39.06	 & 10902 \\
    9        & 3 (QP: 32) & 36.56	 & 5181 \\
    9        & 4 (QP: 37) & 33.97	 & 2678 \\
	\midrule
    10       & 1 (QP: 22) & 41.60	 & 28006 \\
    10       & 2 (QP: 27) & 39.00	 & 11378 \\
    10       & 3 (QP: 32) & 36.47	 & 5364 \\
    10       & 4 (QP: 37) & 33.89	 & 2735 \\
	\bottomrule[1.0pt]
	\end{tabular}
}
\end{table}

\subsection{Results}

With the given numerical information in Section~\ref{sec:3-2}, the performance of stochastic streaming is evaluated and compared with following two streaming algorithms:
\begin{itemize}
\item Queue-independent streaming with maximum quality (QP is set to 22), named to \textsf{[QP22]}
\item Queue-independent streaming with minimum quality (QP is set to 37), named to \textsf{[QP37]}
\end{itemize}

The simulation is with following two criteria: (i) various $K$ setting (refer to Section~\ref{sec:3-3-1}) and (ii) various $V$ setting (refer to Section~\ref{sec:3-3-2}).

\begin{figure*}[t!]
	\centering
	\subfigure[$K = 10$]{
		\includegraphics[width=0.95\columnwidth]{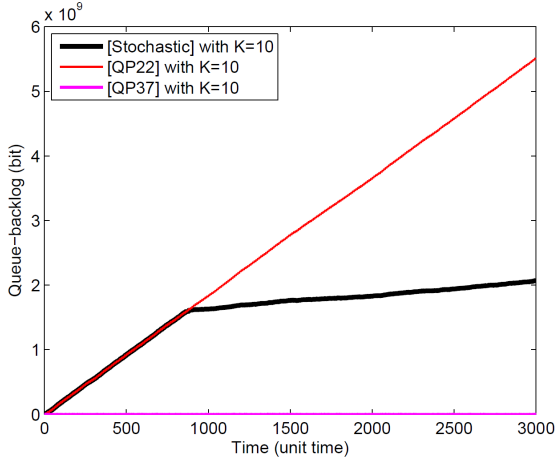}
		\label{fig:simK10}
	}
	\subfigure[$K = 1$]{
		\includegraphics[width=0.95\columnwidth]{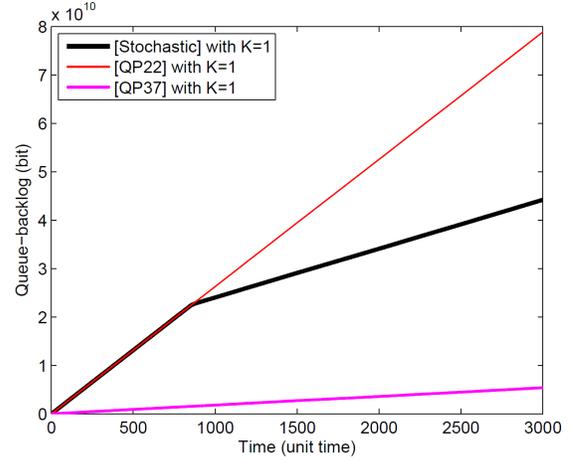}
		\label{fig:simK01}
	}
	\caption{Simulation results with various $K$ ($V=10^{-16}$)}
	\label{fig:simK}
\end{figure*}

\subsubsection{Simulation with various $K$ settings}\label{sec:3-3-1}

For the simulation with various $K$, we consider two cases, i.e., (i) $K=10$ and (ii) $K=1$.
We run the simulation for $3000$ unit times.
In addition, the transmit power and channel bandwidth are assumed to be 5\,dBm and 1\,MHz, respectively.
The simulation results with $K=10$ and $K=1$ are plotted as shown in Fig.~\ref{fig:simK10} and Fig.~\ref{fig:simK01}, respectively.

As shown in both Fig.~\ref{fig:simK10} ($K=10$) and Fig.~\ref{fig:simK01} ($K=1$),
    the queue/buffer diverges if we place streams with the highest quality (i.e., QP=22) as shown in the \textsf{[QP22]} plots.
Fig.~\ref{fig:simK10} shows that the queue/buffer backlog size is always zero if we place streams with the lowest quality (i.e., QP=37).
If $K=1$ (i.e., Fig.~\ref{fig:simK01}), the \textsf{[QP37]} is starting to increase because the placement of streams is frequently occurring.
As shown in both Fig.~\ref{fig:simK10} ($K=10$) and Fig.~\ref{fig:simK01} ($K=1$),
    the proposed stochastic streaming starts to show convergence trends when the unit time is approximately $800$.
Even though both shows convergence trends, the case with $K=10$ is more stable because it places the streams sparser than the placement of streams with $K=1$.

\begin{figure}[t!]
	\begin{center}
		\includegraphics[width=0.95\columnwidth]{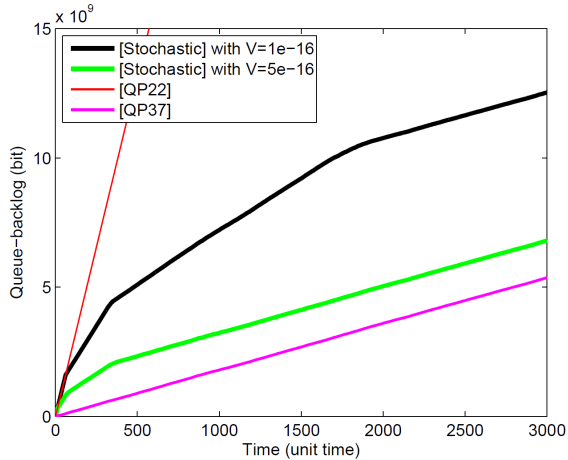}
	\end{center}
	\caption{Simulation results with various $V$ ($K=1$)}
	\label{fig:simV}
\end{figure}

\subsubsection{Simulation with various $V$ settings}\label{sec:3-3-2}

For the simulation with various $V$, we consider two cases, i.e., (i) $V=10^{-16}$ and (ii) $V=5\times 10^{-16}$.
We run the simulation for $3000$ unit times and we also assume that $K=1$.
In addition, the transmit power and channel bandwidth are assumed to be 5\,dBm and 1\,MHz, respectively.
The simulation results with $V=10^{-16}$ and $V=5\times 10^{-16}$ are plotted as shown in Fig.~\ref{fig:simV}.

As shown in Fig.~\ref{fig:simV},
    the queue/buffer also diverges if we place streams with the highest quality (i.e., QP=22) as shown in the \textsf{[QP22]} plots.
As presented in Fig.~\ref{fig:simV},
    both the stochastic streaming algorithms with $V=10^{-16}$ and $V=5\times 10^{-16}$ also present convergence trends; and the algorithm with $V=10^{-16}$ shows the higher queue-backlog sizes because the algorithm with lower $V$ pursues time-average quality maximization rather than queue stability.

\section{Conclusions and Future Work}\label{sec:4}
This paper shows the feasibility study results of stochastic streaming algorithms with 4K ultra-high-definition (UHD) video traces.
In literatures, various stochastic streaming algorithms have been proposed that maximize time-average streaming quality subject to queue stability under the consideration of queue-backlog size.
The performance improvements with the stochastic video streaming algorithms were verified with traditional MPEG test sequences in previous work; however there are no research results with up-to-date 4K UHD video traces.
Thus, this paper verifies the performance of the stochastic streaming algorithms with 4K UHD video traces; and shows that the stochastic algorithms perform better than queue-independent algorithms.

\section*{Acknowledgement}
This research was supported by Basic Science Research Program through the National Research Foundation of Korea(NRF) funded by the Ministry of Science, ICT \& Future Planning (NRF-2015R1C1A1A02037743).
E.-S. Ryu is a corresponding author of this paper.

\end{document}